\documentstyle[preprint,revtex]{aps}
\math-with-secnums
\begin{document}
\bibliographystyle{plain}
\hskip -2\wd0\copy1
\centerline{\bf
Phase diagram of the one-dimensional extended}
\centerline{\bf
Hubbard model at quarter filling
}
\vskip0.5truecm
\author {Fr\'ed\'eric Mila$^{(a)}$ and Xenophon Zotos$^{(b)}$}
\vskip0.5truecm
\begin{instit}
     $(a)$ Institut de Physique, Universit\'e de Neuch\^atel\\
     1 Rue Breguet, CH-2000 Neuch\^atel (Switzerland)\\
$(b)$ Institut Romand de Recherche Num\'erique en Physique des Mat\'eriaux\\
     PHB-Ecublens, CH-1015 Lausanne (Switzerland)
\end{instit}

\begin{abstract}
We used exact diagonalization of small clusters and exact results in various
limits to determine the phase diagram
and the critical exponents of the one dimensional (1D) $U-V$ model at
quarter-filling. We found an instability
of the Luttinger liquid to a charge-density wave (CDW)
insulator
across a boundary going from ($U$,$V$) =($+\infty$,$2t$) to
($4t$,$+\infty$). In the metallic phase, the
dominant fluctuations are superconducting if $V$ is large enough and
spin-density wave (SDW) otherwise. In the latter case,
the critical exponent $\alpha$ of the momentum distribution does not exceed
9/16. The relevance of these results to 1D organic conductors is discussed.
\end{abstract}
\vskip 0.1 truein

\vskip 0.1 truein
\noindent January 1993.
\newpage

The standard description of the electronic properties of quasi 1D compounds is
based on a phenomenological description of the interactions known as
$g$-ology\cite{soly,emer,firs}. This description is phenomenological in the
sense that the $g$-parameters are not directly related to the microscopic
properties of the materials, but are effective parameters intended to describe
the low-energy, long wavelength physics of the system. Once $g$-parameters have
been determined that could explain the low energy data, one must still answer
the question: Can we find a microscopic description of the system which is
reasonable (i.e. compatible with its high energy properties) and which yields
the correct $g$-parameters? This step is absolutely necessary to  get a full
understanding of the materials under study, but it can be very difficult. In
fact, one can go from a microscopic model to the $g$-ology description in a
controlled, analytic way in two cases only: If the  interaction terms are
small,
or if the model has an exact solution (in most cases of the Bethe ansatz form).
Hopelessly, these two cases are far from covering the full range of
$g$-parameters. A very important example is the case of the organic conductors
of the (TMTSF)$_2$X family. In a large temperature range, they exhibit
properties that have a clear one-dimensional character and that are compatible
with the Luttinger liquid picture\cite{wzie,dard,schu} if the exponent $\alpha$
defined by $n(k)-n(k_F) \propto {\rm sign}(k_F-k)\vert k-k_F |^\alpha$ is
slightly larger than $1$, where $n(k)$ is the momentum distribution function.
This value lies far beyond the reach of the above-mentioned methods: Small
interactions always yield $\alpha\ll 1$, while for the two prominent examples
of
strongly correlated models that are soluble by Bethe ansatz (Hubbard\cite{lieb}
and $t-J$\cite{suth,schl,bare} models) $\alpha\leq  1/8$. So, one clearly has
to
turn to other models to have a chance to get large values of $\alpha$.

In this paper, we analyze the possibility to get such a large exponent within
the context of purely electronic interactions. Such systems are characterized
by
the band-filling $n$ and by the form of the interaction term. In the
temperature
range where one-dimensional properties have been reported for (TMTSF)$_2$X, the
dimerization is believed to be unimportant, and the band is
$3/4$-filled\cite{firs}. By electron-hole symmetry, this is equivalent to
quarter-filling ($n=1/2$), and we will concentrate on that case. For the
interaction, one has to go beyond on-site repulsion to get $\alpha\geq 1/8$. As
such models do not have in general an exact solution, one has to resort to
approximations, and the choice of a particular form of the interaction should
be
guided by the possibility of making contact with exact results. In that
respect,
by far the best choice is the $U-V$ model defined by
\begin{equation}
H=-t\sum_{i,\sigma}(c{}^\dagger_{i\sigma}c{}^{\ }_{i+1\sigma}+h.c.)
+U\sum_i n_{i\uparrow}n_{i\downarrow}
+V\sum_i n_in_{i+1}
\label{1}
\end{equation}
because exact results can be obtained in three limits ($V=0$, $U=+\infty$,
$V=+\infty$). Note that this model is also quite reasonable from a physical
point of view, $U$ and $V$ being a priori the largest coupling constants. So,
we
will concentrate on the analysis of Hamiltonian (\ref{1}) at quarter-filling in
the rest of this paper. Some remarks on the possible effect of further
coupling are given at the end.

Let us start by a short summary of what is already known about this model.
For $V=0$, we recover the regular Hubbard model.
At quarter-filling, it is known to be metallic\cite{lieb}, and its low-energy
excitations are of the Luttinger liquid\cite{hald} type with an exponent
$\alpha$ that goes from 0 ($U=0$) to $1/8$ ($U=+\infty$). For $U=+\infty$, we
learn from the equivalence of the charge degrees of freedom to spinless
fermions\cite{schu} that, upon  increasing $V$, the quarter-filled system
undergoes a metal-insulator (M-I) transition
from a Luttinger liquid phase to a CDW insulator at $V=2t$.
The critical exponents for spinless fermions have been calculated by
Haldane\cite{hald}, and  from them one can deduce that $\alpha$ increases from
1/8 at $V=0$ to 9/16 at $V=2t$.
These results suggest two things about the general case: first, the system
is probably a Luttinger liquid for any $U$ as long as $V$ is not too large;
second, there should be an insulating phase with a boundary terminating at
($U$,$V$)=($+\infty$, $2t$).

Before we can say anything about the Luttinger liquid exponents, we have to
know
the range of  parameters for which the system is metallic. So, our first task
is
to determine the boundary of the insulating region. This can be done by a
finite-size analysis of the gap function $\Delta(L;N)$ defined by
\begin{equation}
\Delta(L;N)=E_0(L;N+1)+E_0(L;N-1)-2E_0(L;N)
\label{2}
\end{equation}
where $E_0(L;M)$, the ground-state energy of $M$  particles on $L$ sites, can
be
obtained by exact diagonalization of small clusters using Lanczos algorithm.
Such a procedure has been used by Spronken et al\cite{spro} for the case of
spinless fermions with nearest neighbour repulsion,
and the critical value $V_c=2t$ could be reproduced quite accurately by fitting
$\Delta(L;N)$ with polynomials of $1/L$. In our case,
the Hilbert space is much larger, and we have results
for 3 sizes only ($L$=8,12,16). So, we cannot expect a very good accuracy. In
fact, we could not get meaningful results by fitting these 3 values with
$A+B/L+C/L^2$. However, fitting any pair of values with $A+B/L$ gives
reasonable results that do not depend too much on the pair chosen to do the
fit. Besides, as far as the large $U$ case is concerned, the best result is
obtained by fitting the results for $L=8$ and $16$.
This is probably due to the fact that the $12$ site system is quite different
from the other two: To get smooth results as a function of $L$, one must
choose the boundary conditions so that $k_F$ is one of the allowed $k$ values,
that is periodic ones for 8 and 16 sites, and antiperiodic
ones for 12 sites\cite{spro}.
The points obtained with $L$=8 and 16 are given in figure 1 (open circles).

In view of the uncertainties of the finite-size analysis, one would like to
have independent information on the boundary. As we shall see below,
one can check that the system is a Luttinger liquid by estimating the central
charge or with the help of sum rules.
An even better check would be to know another point besides
($U$,$V$)=($+\infty$,$2t$)
that lies on the boundary. It turns out that the exact value of $U$ for
$V=+\infty$ can be obtained quite easily, and we now turn to a discussion of
this limit. When $V=+\infty$, all the configurations having two particles on
neighbouring sites must be excluded from the Hilbert space. The essential
consequence is that the number of local pairs (i.e. pairs of particles on the
same site) is conserved: to break a pair (or to make a pair out of
two particles that are far apart) one should overcome an infinite energy
barrier. For the same reason, these pairs cannot move. So, we can classify the
states according to the number and position of the pairs. The remaining
particles are confined to intervals between pairs, and the number of unpaired
particles  in a given interval is again conserved. As these unpaired
particles cannot cross, there is a full spin degeneracy, and the energy levels
are the same as for spinless fermions with an infinite repulsion between
nearest neighbours, a problem that can be solved by Bethe
ansatz\cite{yang,fowl}. This allows
one to get the full spectrum of Hamiltonian (1) in the $V=+\infty$ limit. A
detailed discussion will be given elsewhere. For our present purpose, all we
need is an evaluation of the gap $\Delta(L=2N;N)$ when $L\rightarrow +\infty$.
For $U$ large enough, one can easily show that
the ground-state minimizes the number of pairs. With
$N=L/2$ particles, the system then has no pair but a single particle every
two sites, and it will not be able to gain any kinetic energy. Hence,
$E_0(L;N)=0$. The system with $N+1$ particles is forced to have one pair, with
a cost $U$ in energy, and it cannot gain any kinetic energy because the pair
cannot move.
So, $E_0(L;N+1)=U$. Finally, the system with $N-1$ particles has no pair and
can gain
kinetic energy. The simplest way to think about it is to remove a particle from
the state with $N$ particles. One then has 3 empty sites side by side. The
empty sites at the left and right sides of this trimer can then move
independently\cite{fowl},
and in the limit $L\rightarrow +\infty$, each of them will gain $2t$, so that
$E_0(L;N-1)=-4t$. This result can of course be obtained rigorously from the
Bethe ansatz equations.
Finally, $\lim_{N\rightarrow+\infty}\Delta(L=2N;N)=U-4t$ is
positive for large $U$ and vanishes for $U_c=4t$, which is the critical value
we
were looking for in the limit $V=+\infty$. Our numerical results are in good
agreement with this value. It seems that the boundary is horizontal between
$V=4t$ and $V=+\infty$, although a weak dependence on $V$ cannot be excluded.

We now turn to the analysis of the metallic phase. In the weak coupling limit,
the model can be studied with the help of $g$-ology\cite{soly,emer}. At
quarter-filling, the
coupling constants are $g_1=U$, $g_2=U+2V$, $g_{4\parallel}=2V$ and
$g_{4\perp}=U+2V$. For $U>0$, the system scales to the Tomonaga-Luttinger fixed
point $g_1^*=0$, $g_2^*=g_2-g_1/2$. So, the long wavelength properties depend
on 3 parameters only. One usually chooses the velocities of the charge (resp.
spin) density oscillations $u_\rho$ (resp. $u_\sigma$) and a parameter $K_\rho$
that determines the exponent of the correlation functions. In particular, the
exponent $\alpha$ is given by $\alpha=(K_\rho+1/K_\rho-2)/4$.

In the strong coupling limit,
perturbation theory cannot be invoked to prove  that the long-wavelength
physics
can be described by the Tomonaga-Luttinger fixed point. However, assuming this
to be the case, one can deduce the parameters of the model from the low energy
part of the spectrum\cite{hald,frah} (which can be obtained numerically for
small clusters, as suggested by Schulz\cite{schu}) and check a posteriori if
these parameters are consistent with the hypothesis. One can estimate the ratio
$u_\rho/K_\rho$ by using its relation to the compressibility $\kappa$
\begin{equation}
{\pi \over 2} {u_\rho \over K_\rho} =
{1\over n^2\kappa}
\label{3 a}
\end{equation}
\begin{equation}
\kappa={L\over N^2} \biggl({E_0(N+2)+E_0(N-2)-2E_0(N) \over 4}\biggr)^{-1}
\label{3 b}
\end{equation}
Equation (\ref{3 b}) is the
finite-size approximation to the compressibility,
$E_0(N)$ being the ground-state energy calculated with suitable
boundary conditions. The
charge velocity can be obtained from the low-energy spectrum as
\begin{equation}
u_\rho=(E_{1\rho}-E_0)/(2\pi/L)
\label{4}
\end{equation}
The analysis of the $t-J$ model by Ogata et al\cite{ogat} was based on these
equations.

In our study, we have also used another relation that holds for Luttinger
liquid, namely
\begin{equation}
\sigma_0=2u_\rho K_\rho
\label{5 a}
\end{equation}
where $\sigma_0$ is the weight of the Drude peak of the conductivity.
In 1D systems, $\sigma_0$ can be obtained simply from\cite{kohn,zoto,shas}
\begin{equation}
\sigma_0={\pi \over L} {\partial^2 E_0(\phi) \over
\partial \phi^2} \Biggr |_{\phi=0}
\label{5 b}
\end{equation}

\noindent where $E_0(\phi)$ is the ground-state energy as a function of
a phase $\phi$ due to the flux through the ring.
Equations (\ref{3 a} -\ref{5 b}) provide us with 3
independent conditions on $u_\rho$ and $K_\rho$. This is very useful for
several
reasons. First, the conductivity is much simpler to evaluate numerically than
the compressibility, which involves systems with $N+2$ particles.  Besides, we
have good reasons to believe that equation (\ref{3 b}) does not give a reliable
estimate of the compressibility when $V$ is large, and it is important to be
able to determine $K_\rho$ without its help.

We can use the consistency of these equations to
check the assumption that the system is a Luttinger liquid:
Equations (\ref{3 a}-\ref{3 b}) and (\ref{5 a}-\ref{5 b}) are typical of
Luttinger liquids and will be violated if we have an
instability to another phase (e.g. CDW insulator). A convenient way to measure
the consistency of the three conditions is to calculate the ratio
$\sigma_0 / \pi n^2 \kappa u^2_\rho$
which equals 1 for a Luttinger liquid. We have performed a systematic
evaluation of this ratio along lines in the ($U,V$) plane.
In the region where equation (\ref{4}) can be used, i.e. when $V$ is not too
large,
we found that, to a
good accuracy, this ratio equals 1 in the metallic phase, that it
drops rapidly when one enters the insulating phase, and that the transition
point was in good agreement with our previous determination of the phase
boundary.

Another way to check that we have a Luttinger liquid is to extract the central
charge from the finite-size scaling of the ground-state energy density
\cite{frah}
\begin{equation}
e_0(L)=e_0(+\infty)-{\pi(u_\rho + u_\sigma) \over 6 L^2} c +o(1/L^2)
\label{7}
\end{equation}
$u_\sigma$ was obtained from $u_\sigma=(E_{1\sigma} -E_0)/(2\pi/L)$,
where $E_{1\sigma}$ is the energy of the first excited state with total spin
$S\ne 0$. Comparing our results for $L$=8, 12 and 16, we found that this
scaling
form was accurate except when $V$ is too large, and that the
central charge $c$ determined in this way equals 1 within $2\%$, in good
agreement with the exact value $c=1$ for Luttinger liquids.

Let us next discuss the exponents in this Luttinger liquid phase. We have
performed a systematic determination of $u_\rho$ and $K_\rho$ with the help of
equations (\ref{4}-\ref{5 b}) for clusters of 8, 12 and 16 sites by Lanczos
diagonalization.  The constant $K_\rho$ curves obtained for 16 sites are
plotted
in figure 1 (dashed curves). When $V$ is not too large, a meaningful
extrapolation to the thermodynamic limit is possible, and the values obtained
for $L$=16 are quite close to the extrapolated ones. This corresponds to the
short-dashed curves. However, when $V$ gets large, various finite-size effects
are present (see below), and the curves for $K_\rho \ge 0.8$ (long-dashed
lines)
are only indicative.

The qualitative features
that emerge are the following. As long as $V$ is not too large, $K_\rho$
decreases when $V$ increases. So, the exponent $\alpha$ effectively increases
with $V$. The largest values are obtained close to the CDW boundary and do not
exceed say 0.6.
For $U<4t$, $K_\rho$ is non monotonic and starts increasing when
$V$ is large enough. This results into a region where $K_\rho>1$, i.e. where
superconducting fluctuations are dominant.
Some physical insight about the nature of this phase can be gained by looking
again at the $V=+\infty$ case. We saw that the transition between the CDW phase
and the metallic phase corresponds to the formation of a local pair. When $U$
decreases further, more and more local pairs are formed. In the thermodynamic
limit, the Bethe ansatz results\cite{fowl} can be used to show that the density
$n_p$ of paired particles is given by
\begin{equation}
n_p=(1/ 2\pi) \arccos(U/4)
\label{8}
\end{equation}
In the limit $U=0$, $n_p=1/4$, i.e. half the particles are involved in pairs.
When $V$ is large but finite, these pairs are probably mobile and yield strong
superconducting fluctuations. Let us note however that a full understanding of
this region requires a more careful analysis.  The first source of concern is
the reliability of finite-size results in this region. For $V=+\infty$, the
formation of pairs is strongly affected by the size of the system. For
instance,
the critical value where the first pair forms is $U_c$ = $2.236\ t$, $3.049\ t$
and $3.411\ t$ for $L$=8, 12 and 16, to be compared with $U_c=4\ t$ for the
infinite system. Besides, the $L$=12 system has 6 particles and cannot have
half
its particles paired.  For the $L$=16  system, the second pair is formed for
$U=1.347\ t$. This abrupt change in the density of unpaired particles is
responsible for irregularities in the constant $K_\rho$ curves. These
irregularities will disappear in the thermodynamic limit, and we have preferred
to give a linear fit to the results (long-dashed lines). Nevertheless, a region
where $K_\rho>1$ was found for all sizes for roughly the same parameters, and
the existence of such a region in the thermodynamic limit is very likely.

The second source of concern about this superconducting phase comes from the
competition with phase separation, a phenomenon already observed for instance
for the $t-J$ model\cite{ogat}. For the present model, it can be
shown\cite{penc} that the ground-state exhibits phase separation when
$V=+\infty$. Preliminary results about the compressibility indicate that this
phase separation occurs beyond the $K_\rho=1$ line, and that a region of
divergent superconducting fluctuations survives. However, due to the various
finite-size effects outlined above, no definite answer can be given on the
basis
of the available numerical results, and we are currently working on new
approaches to the problem.

Finally, coming back to the original problem of finding a microscopic model
with
$\alpha$ slightly larger than 1 at quarter-filling, we have shown that this is
not possible within the model of equation (\ref{1}). Large values of $\alpha$,
that are possible away from quarter-filling\cite{hald,schu}, cannot be obtained
when $n=1/2$ because of the M-I transition. This does not exclude the
possibility that other models with longer range interactions can do the job.
However, a remarkable property of our results suggests that this will not be
the
case: The boundary between the insulating and the metallic phases corresponds
within the accuracy of our results to the curve $\alpha$ maximum. So, one may
conjecture that the M-I transition is related to the value of $\alpha$, in
which case other models will {\it not} be able to produce larger values of
$\alpha$ before the insulating phase is reached.
This property is consistent with the observation  that the $4k_F$
fluctuations\cite{emer2,lee} become divergent along a line $\alpha$ constant
($\alpha=1/8$).
The M-I transition, i.e. the opening of a gap in
the charge-excitation spectrum, is a different effect however which, according
to our results, occurs at a considerably larger value of $\alpha$ (9/16).
So far as we know, a weak-coupling theory of commensurability effects
has been worked out for half-filling only\cite{nogu}, and more work
is needed to confirm or
dismiss our conjecture about the quarter-filled case. We are currently working
on this problem.

We acknowledge useful discussions with D. Baeriswyl, H. Beck, K. Penc,
T. M. Rice and the group of Y. Baer.
We are particularly indebted to D. Poilblanc for fruitful suggestions
about the program and to H. Schulz for his interest in this project. This work
was supported in part by the Swiss National Science Foundation under Grants
Nos. 20-30272.90 and 20-31125.91 and by the University of Fribourg.

\figure{ Phase diagram of $U-V$ model at quarter-filling. M-metal,
I-insulating,
SC-superconducting phases. Solid line and open circles: M-I transition; arrows:
exact limiting values of the M-I boundary; dashed lines and full circles:
constant $K_\rho$ curves.}
\end{document}